\documentclass[11pt,a4paper]{article}
\usepackage[hyperref]{acl2018}
\usepackage{times}
\usepackage{latexsym}
\usepackage{url}
\usepackage{color}
\usepackage{enumitem}

\aclfinalcopy %
\def\aclpaperid{969} %

\usepackage{multirow}
\usepackage{subfig}
\usepackage{amsmath,amssymb}
\DeclareMathOperator*{\argmin}{arg\,min}
\DeclareMathOperator*{\argmax}{arg\,max}
\usepackage{algorithm}
\usepackage{algorithmic}
\usepackage{graphicx}

\newcommand*{\affmark}[1][*]{\textsuperscript{#1}}
\newcommand*{\affaddr}[1]{#1}

\title{A Purely End-to-end System for Multi-speaker Speech Recognition}

\author{Hiroshi Seki\affmark[1,2]\thanks{This work was done while H. Seki, Ph.D.\ candidate at Toyohashi University of Technology, Japan, was an intern at MERL.}, Takaaki Hori\affmark[1], Shinji Watanabe\affmark[3], Jonathan Le Roux\affmark[1], John R.\ Hershey\affmark[1] \\
\affaddr{\affmark[1]Mitsubishi Electric Research Laboratories (MERL)} \\
\affaddr{\affmark[2]Toyohashi University of Technology} \\
\affaddr{\affmark[3]Johns Hopkins University}
}

\date{}

\begin{document}
\maketitle
\begin{abstract} %
Recently, there has been growing interest in multi-speaker speech recognition, where the utterances of multiple speakers are recognized from their mixture. Promising techniques have been proposed for this task, but earlier works have required additional training data such as isolated source signals or senone alignments for effective learning. In this paper, we propose a new sequence-to-sequence framework to directly decode multiple label sequences from a single speech sequence by unifying source separation and speech recognition functions in an end-to-end manner. We further propose a new objective function to improve the contrast between the hidden vectors to avoid generating similar hypotheses. Experimental results show that the model is directly able to learn a mapping from a speech mixture to multiple label sequences, achieving 83.1\% relative improvement compared to a model trained without the proposed objective. Interestingly, the results are comparable to those produced by previous end-to-end works featuring explicit separation and recognition modules. 
\end{abstract}

\section{Introduction}
\label{seq:intro}
Conventional automatic speech recognition (ASR) systems recognize a single utterance given a speech signal, in a one-to-one transformation.
However, restricting the use of ASR systems to situations with only a single speaker limits their applicability.
Recently, there has been growing interest in single-channel multi-speaker speech recognition, 
which aims at generating multiple transcriptions from a single-channel mixture of multiple speakers' speech~\cite{Cooke2009monaural}.

To achieve this goal, several previous works have considered a two-step procedure in which the mixed speech is first separated, and recognition is then performed on each separated speech signal~\cite{john2016deep, Isik2016, dong2017permutation, chen2017deep}.
Dramatic advances have recently been made in speech separation,  via the \emph{deep clustering} framework~\cite{john2016deep, Isik2016}, hereafter referred to as DPCL. DPCL trains a deep neural network to map each time-frequency (T-F) unit to a high-dimensional embedding vector such that the embeddings for the T-F unit pairs dominated by the same speaker are close to each other, while those for pairs dominated by different speakers are farther away. The speaker assignment of each T-F unit can thus be inferred from the embeddings by simple clustering algorithms, to produce masks that isolate each speaker. The original method using k-means clustering \cite{john2016deep} was extended to allow end-to-end training by unfolding the clustering steps using a permutation-free mask inference objective \cite{Isik2016}. An alternative approach is to  perform \emph{direct mask inference} using the permutation-free objective function with networks that directly estimate the labels for a fixed number of sources.  Direct mask inference was first used  in Hershey et al.~\shortcite{john2016deep} as a baseline method, %
but without showing good performance. %
This approach was revisited in Yu et al.~\shortcite{dong2017permutation} and Kolbaek et al.~\shortcite{kolbaek2017multitalker} under the name permutation-invariant training (PIT). %
Combination of such single-channel speaker-independent multi-speaker speech separation systems with ASR was first considered in Isik et al.~\shortcite{Isik2016} using a conventional Gaussian Mixture Model/Hidden Markov Model (GMM/HMM) system. 
Combination with an end-to-end ASR system was recently proposed in \cite{Settle2018ICASSP04}.
Both these approaches either trained or pre-trained the source separation and ASR networks separately, making use of mixtures and their corresponding isolated clean source references. While the latter approach could in principle be trained without references for the isolated speech signals, the authors found it difficult to train from scratch in that case. This ability can nonetheless be used when adapting a pre-trained network to new data without such references.

In contrast with this two-stage approach, Qian et al.\ \shortcite{Qian2017single} considered direct optimization of a deep-learning-based ASR recognizer without an explicit separation module. The network is optimized based on a permutation-free objective defined using the cross-entropy between the system's hypotheses and reference labels. The best permutation between hypotheses and reference labels in terms of cross-entropy is selected and used for backpropagation. However, this method still requires reference labels in the form of senone alignments, which have to be obtained on the clean isolated sources using a single-speaker ASR system. As a result, this approach still requires the original separated sources.
As a general caveat, generation of multiple hypotheses in such a system requires the  number of speakers handled by the neural network architecture to be determined before training. %
However, Qian et al.~\shortcite{Qian2017single} reported that the recognition of two-speaker mixtures using a model trained for three-speaker mixtures showed almost identical performance with that of a model trained on two-speaker mixtures. Therefore, it may be possible in practice to  determine an upper bound  on the number of speakers.

Chen et al.~\shortcite{chen2018progressive} proposed a progressive training procedure for a hybrid system with explicit separation motivated by curriculum learning. They also proposed self-transfer learning and multi-output sequence discriminative training methods for fully exploiting pairwise speech and preventing competing hypotheses, respectively.

In this paper, we propose to circumvent the need for the corresponding isolated speech sources when training on a set of mixtures, by using an end-to-end multi-speaker speech recognition without an explicit speech separation stage.  In separation based systems, the spectrogram is segmented into complementary regions according to sources, which generally ensures that different utterances are recognized for each speaker.  Without this complementarity constraint, our direct multi-speaker recognition system could be susceptible to redundant recognition of the same utterance. In order to prevent degenerate solutions in which the generated hypotheses are similar to each other, we introduce a new objective function that enhances contrast between the network's representations of each source. 
We also propose a training procedure to provide permutation  invariance with low computational cost, by taking advantage of the joint CTC/attention-based encoder-decoder network architecture proposed in \cite{hori-ACL-2017}.
Experimental results show that the proposed model is able to directly convert an input speech mixture into multiple label sequences
without requiring any explicit intermediate representations.  In particular no frame-level  training labels, such as  phonetic alignments  or corresponding unmixed speech, are required. %
We evaluate our model on spontaneous English and Japanese tasks and obtain comparable results to the DPCL based method with explicit separation \cite{Settle2018ICASSP04}.

\section{Single-speaker  end-to-end ASR}
\label{sec:singleasr}
\subsection{Attention-based encoder-decoder network}
An attention-based encoder-decoder network~\cite{bahdanau2016end} predicts a target label sequence $Y = (y_1, \ldots, y_N)$ without requiring intermediate representation from a $T$-frame sequence of $D$-dimensional input feature vectors, $O = ( o_t \in \mathbb{R}^D | t = 1, \dots, T)$, and the past label history.
The probability of the $n$-th label $y_n$ is computed by conditioning on the past history $y_{1:n-1}$:
\begin{align}
    p_\text{att}(Y|O) &= \prod_{n=1}^N p_\text{att}(y_{n}|O, y_{1:n-1}). \label{eq:att_final}
\end{align}
The model is composed of two main sub-modules, an encoder network and a decoder network.
The encoder network %
transforms the input feature vector sequence into a high-level representation $H = ( h_l \in \mathbb{R}^C | l = 1, \dots, L)$.
The decoder network %
emits labels based on the label history $y$ and a context vector $c$ calculated using an attention mechanism which weights and sums the $C$-dimensional sequence of representation $H$ with attention weight $a$.
A hidden state $e$ of the decoder is updated based on the previous state, the previous context vector, and the emitted label. This mechanism is summarized as follows:
\begin{align}
    H &= \mathrm{Encoder}(O), \label{eq:enc} \\
    y_{n} & \sim \mathrm{Decoder}(c_{n}, y_{n-1}),  \label{eq:dec} \\
    c_{n},a_{n} &= \mathrm{Attention}(a_{n-1}, e_{n}, H), \label{eq:att} \\
    e_{n} &= \mathrm{Update}(e_{n-1}, c_{n-1}, y_{n-1}). \label{eq:dec_begin}
\end{align}
At inference time, the previously emitted labels are used. At training time, they are replaced by the reference label sequence $R = (r_1, \ldots, r_{N})$ in a \emph{teacher-forcing} fashion, leading to conditional probability $p_\text{att}(Y_{R}|O)$, where $Y_{R}$ denotes the output label sequence variable in this condition. 
The detailed definitions of $\mathrm{Attention}$ and $\mathrm{Update}$ are described in Section~\ref{sup:encdec} of the supplementary material. 
The encoder and decoder networks are trained to maximize the conditional probability of the reference label sequence $R$ using backpropagation:
\begin{align}
    \mathcal{L}_\text{att} &= \mathrm{Loss}_\text{att}(Y_{R}, R) \triangleq -\log p_\text{att}(Y_{R}=R|O),
\end{align}
where $\mathrm{Loss}_\text{att}$ is the cross-entropy loss function.

\subsection{Joint CTC/attention-based encoder-decoder network}
\label{sec:jointctcatt}
The joint CTC/attention approach \cite{kim2016joint_icassp2017,hori-ACL-2017},
uses the connectionist temporal classification (CTC) objective function~\cite{graves2006connectionist} as an auxiliary task to train the network.
CTC formulates the conditional probability by introducing a framewise label sequence $Z$
consisting of a label set $\mathcal{U}$ and an additional blank symbol
defined as $Z=\{ z_l \in \mathcal{U} \cup \{\texttt{'blank'}\} | l=1,\cdots,L \}$:
\begin{align}
p_\text{ctc}(Y|O) & =
 \sum _{Z} \prod _{l=1} ^L p(z_l | z _{l-1}, Y) p(z_l|O) \label{eq:ctc_final},
\end{align}
where $p(z_l|z_{l-1}, Y)$ represents monotonic alignment constraints in CTC and $p(z_l|O)$ is the frame-level label probability computed by
\begin{align}
p(z_l|O) & = \mathrm{Softmax}(\mathrm{Linear}(h_l)),
\label{eq:ctc_frame}
\end{align}
where $h_l$ is the hidden representation generated by an encoder network, here taken to be the encoder of the attention-based encoder-decoder network defined in Eq.~(\ref{eq:enc}), and $\mathrm{Linear}(\cdot)$ is the final linear layer of the CTC to match the number of labels. 
Unlike the attention model, the forward-backward algorithm of CTC enforces monotonic alignment 
between the input speech and the output label sequences during training and decoding.
We adopt the joint CTC/attention-based encoder-decoder network as the monotonic alignment helps the separation and extraction of high-level representation.
The CTC loss is calculated as:
\begin{align}
    \mathcal{L}_\text{ctc} = \mathrm{Loss}_\text{ctc}(Y, R) \triangleq -\log p_\text{ctc}(Y=R|O).
\end{align}
The CTC loss and the attention-based encoder-decoder loss are combined with an interpolation weight $\lambda \in [0,1]$:
\begin{align}
    \mathcal{L}_\text{mtl} = \lambda \mathcal{L}_\text{ctc} + (1-\lambda) \mathcal{L}_\text{att}. \label{eq:single_spk_mtl}
\end{align}

Both CTC and encoder-decoder networks are also used in the inference step. The final hypothesis is a sequence that maximizes a weighted conditional probability of CTC in Eq.~(~\ref{eq:ctc_final}) and attention-based encoder decoder network in Eq.~(\ref{eq:att_final}):
\begin{multline}
    \hat{Y} = \argmax_{Y} \bigl\{ \gamma \log p_\text{ctc}(Y|O) \\ 
    + (1-\gamma) \log p_\text{att} (Y|O) \bigr\} \label{eq:joint_dec},
\end{multline}
where $\gamma \in [0,1]$ is an interpolation weight.

\section{Multi-speaker end-to-end ASR}

\subsection{Permutation-free training}
In situations where the correspondence between the outputs of an algorithm and the references is an arbitrary permutation, neural network training faces a {\it permutation problem}. This problem was first addressed by deep clustering \cite{john2016deep}, which circumvented it in the case of source separation by comparing the relationships between pairs of network outputs to those between pairs of labels. As a baseline for deep clustering, Hershey et al.~\shortcite{john2016deep} also proposed another approach to address the permutation problem, based on an objective which considers all permutations of references when computing the error with the network estimates. This objective was later used in Isik et al.~\shortcite{Isik2016} and Yu et al.~\shortcite{dong2017permutation}. In the latter, it was referred to as permutation-invariant training.

This permutation-free training scheme extends the usual one-to-one mapping of outputs and labels for backpropagation to one-to-many by selecting the proper permutation of hypotheses and references,
thus allowing the network to generate multiple independent hypotheses from a single-channel speech mixture.
When a speech mixture contains speech uttered by $S$ speakers simultaneously, the network generates $S$ label sequence variables 
$Y^s = (y_1^s, \ldots, y_{N_s}^s)$ with $N_s$ labels
from the $T$-frame sequence of $D$-dimensional input feature vectors, $O = ( o_t \in \mathbb{R}^D | t = 1, \dots, T)$:
\begin{align}
    Y^s \sim g^s(O), \ s = 1, \ldots, S,
\end{align}
where the transformations $g^s$ are implemented as neural networks which typically share some components with each other.
In the training stage, all possible permutations of the $S$ sequences $R^s = (r_1^s, \ldots, r_{N'_s}^s)$ of $N'_s$ reference labels are considered (considering permutations on the hypotheses would be equivalent), and the one leading to minimum loss is adopted for backpropagation.
Let $\mathcal{P}$ denote the set of permutations on $\{1,\dots,S\}$.
The final loss $\mathcal{L}$ is defined as %
\begin{align}
    \mathcal{L} = \min_{\pi \in \mathcal{P}} \sum_{s=1}^{S} {\mathrm{Loss}}(Y^{s}, R^{\pi(s)}),
\end{align}
where $\pi(s)$ is the $s$-th element of a permutation $\pi$.
For example, for two speakers, $\mathcal{P}$ includes two permutations $(1, 2)$ and $(2, 1$), and the loss is defined as:
\begin{multline}
    \mathcal{L} = \min (\mathrm{Loss}(Y^1,R^1)+\mathrm{Loss}(Y^2,R^2),\\
    \mathrm{Loss}(Y^1,R^2)+\mathrm{Loss}(Y^2,R^1)).
\end{multline}

\begin{figure}[t]
\vskip 0.2in
\begin{center}
\centerline{\includegraphics[width=7cm]{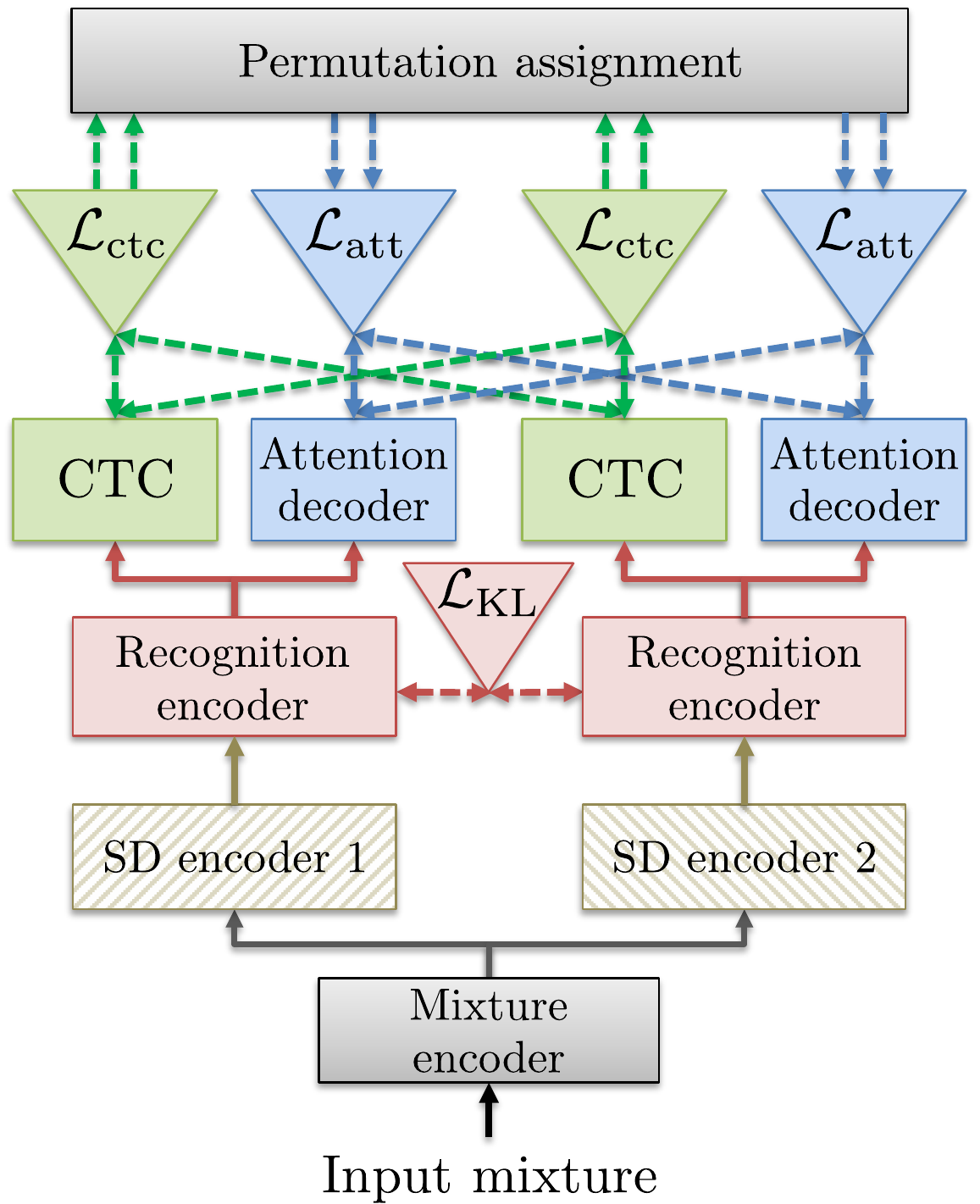}}
\caption{End-to-end multi-speaker speech recognition. 
We propose to use the permutation-free training for CTC and attention loss functions $\mathrm{Loss}_{\text{ctc}}$ and $\mathrm{Loss}_{\text{att}}$, respectively.
}
\label{fig:end-to-end-pft}
\end{center}
\vskip -0.2in
\end{figure}

Figure~\ref{fig:end-to-end-pft} shows an overview of the proposed end-to-end multi-speaker ASR system.
In the following Section~\ref{subsec:e2epft}, we describe an extension of encoder network for the generation of multiple hidden representations. We further introduce a permutation assignment mechanism for reducing the computation cost in Section~\ref{subsec:sync_perm}, and an additional loss function $\mathcal{L}_{KL}$ for promoting the difference between hidden representations in Section~\ref{subsec:neg_kl_loss}.

\label{sec:e2epit}

\subsection{End-to-end permutation-free training}
\label{subsec:e2epft}
To make the network output multiple hypotheses, we consider a stacked architecture that combines both shared and unshared (or specific) neural network modules. %
The particular architecture we consider in this paper splits the encoder network into three stages: the first stage, also referred to as mixture encoder, processes the input mixture and outputs an intermediate feature sequence $H$; that sequence %
is then processed by $S$ independent encoder sub-networks which do not share parameters, also referred to as speaker-differentiating (SD) encoders, leading to $S$ feature sequences $H^s$; at the last stage, each feature sequence $H^s$ is independently processed by the same network, also referred to as recognition encoder, leading to $S$ final high-level representations $G^s$.

Let $u \in \{1\dots,S\}$ denote an output index (corresponding to the transcription of the speech by one of the speakers), and $v \in \{1\dots,S\}$ denote a reference index. Denoting by ${\rm Encoder_{Mix}}$ the mixture encoder, ${\rm Encoder_{SD}^u}$ the $u$-th speaker-differentiating encoder, and ${\rm Encoder_{Rec}}$  the recognition encoder, an input sequence $O$ corresponding to an input mixture can be processed by the encoder network as follows:
\begin{align}
H &= {\rm Encoder_{Mix}}(O), \label{eq:enc_mix} \\
H^{u} &= {\rm Encoder_{SD}^{u}(H)}, \label{eq:enc_sd} \\
G^{u} &= {\rm Encoder_{Rec}}(H^{u}). \label{eq:enc_rec}
\end{align}
The motivation for designing such an architecture can be explained as follows, following analogies with the architectures in \cite{Isik2016} and \cite{Settle2018ICASSP04} where separation and recognition are  performed explicitly in separate steps: the first stage in Eq.~(\ref{eq:enc_mix}) corresponds to a speech separation module which creates embedding vectors that can be used to distinguish between the multiple sources; the speaker-differentiating second stage in Eq.~(\ref{eq:enc_sd}) uses the first stage's output to disentangle each speaker's speech content from the mixture, and prepare it for recognition; the final stage in Eq.~(\ref{eq:enc_rec}) corresponds to an acoustic model that encodes the single-speaker speech for final decoding. %

{\allowdisplaybreaks
The decoder network computes the conditional probabilities for each speaker from the $S$ outputs of the encoder network.
In general, the decoder network uses the reference label $R$ as a history to generate the attention weights during training, in a teacher-forcing fashion.
However, in the above permutation-free training scheme, the reference label to be attributed to a particular output is not determined until the loss function is computed, so we here need to run the attention decoder for all reference labels. %
We thus need to consider the conditional probability of the decoder output variable $Y^{u,v}$ for each output $G^u$ of the encoder network under the assumption that the reference label for that output is $R^v$:
\begin{align}
&p_\text{att}(Y^{u,v}|O) = \prod_{n} p_\text{att}(y^{u,v}_{n}|O, y^{u,v}_{1:n-1}), \label{eq:naiveprob}\\
&c^{u,v}_{n},a^{u,v}_{n} = \mathrm{Attention}(a^{u,v}_{n-1}, e^{u,v}_{n}, G^u),  \\
&e^{u,v}_{n} = \mathrm{Update}(e^{u,v}_{n-1}, c^{u,v}_{n-1}, r^{v}_{n-1}),  \\
&y^{u,v}_{n} \sim \mathrm{Decoder}(c^{u,v}_{n}, r^{v}_{n-1}).
\end{align}
}
The final loss is then calculated by considering all permutations of the reference labels as follows:
\begin{align}
    \mathcal{L}_\text{att} = \min_{\pi \in \mathcal{P}} \sum_s \mathrm{Loss}_\text{att}(Y^{s,\pi(s)}, R^{\pi(s)}).
\end{align}

\subsection{Reduction of permutation cost}
\label{subsec:sync_perm}
In order to reduce the computational cost, we fixed the permutation of the reference labels based on the minimization of the CTC loss alone, and used the same permutation for the attention mechanism as well. 
This is an advantage of using a joint CTC/attention based end-to-end speech recognition. Permutation is performed only for the CTC loss by assuming synchronous output where the permutation is decided by the output of CTC:
\begin{align}
    &\hat{\pi} = \argmin_{\pi \in \mathcal{P}} \sum_s \mathrm{Loss}_\text{ctc}(Y^{s}, R^{\pi(s)}), \label{eq:ctc_loss}
\end{align}
where $Y^u$ is the output sequence variable corresponding to encoder output $G^u$.
{
Attention-based decoding is then performed on the same hidden representations $G^u$, using teacher forcing with the labels determined by the permutation $\hat{\pi}$ that minimizes the CTC loss:
\begin{align*}
    & p_\text{att}(Y^{u,\hat{\pi}(u)}|O) = \prod_{n} p_\text{att}(y^{u,\hat{\pi}(u)}_{n}|O, y^{u,\hat{\pi}(u)}_{1:n-1}), \\
    &c^{u,\hat{\pi}(u)}_{n}\!\! , a^{u,\hat{\pi}(u)}_{n} \! = \! \mathrm{Attention}(a^{u,\hat{\pi}(u)}_{n-1}\!\! , e^{u,\hat{\pi}(u)}_{n}\!\! , G^u),  \\
    &e^{u,\hat{\pi}(u)}_{n} = \mathrm{Update}(e^{u,\hat{\pi}(u)}_{n-1}\!, c^{u,\hat{\pi}(u)}_{n-1}\!, r^{\hat{\pi}(u)}_{n-1}),\\
    &y^{u,\hat{\pi}(u)}_{n} \sim \mathrm{Decoder}(c^{u,\hat{\pi}(u)}_{n}, r^{\hat{\pi}(u)}_{n-1}).
\end{align*}
This corresponds to the ``permutation assignment'' in Fig.~\ref{fig:end-to-end-pft}. In contrast with Eq.~(\ref{eq:naiveprob}), we only need to run the attention-based decoding once for each output $G^u$ of the encoder network.
The final loss is defined as the sum of two objective functions with interpolation $\lambda$:
\begin{align}
    \mathcal{L}_\text{mtl} &= \lambda \mathcal{L}_\text{ctc} + (1-\lambda) \mathcal{L}_\text{att}, \label{eq:multi_spk_mtl} \\
    \mathcal{L}_\text{ctc} &= \sum_s \mathrm{Loss}_\text{ctc}(Y^{s}, R^{\hat{\pi}(s)}) , \label{eq:ctcloss} \\
    \mathcal{L}_\text{att} &= \sum_s {\mathrm{Loss}}_\text{att}(Y^{s,\hat{\pi}(s)}, R^{\hat{\pi}(s)}). \label{eq:attloss} %
\end{align}}

At inference time, because both CTC and attention-based decoding are performed on the same encoder output $G^u$ and should thus pertain to the same speaker, their scores can be incorporated as follows: %
\begin{align}
    \hat{Y}^u &= \argmax_{{Y^u}} \bigl\{ \gamma \log p_\text{ctc}(Y^u|G^u) \nonumber \\
    & \hspace{1cm} + (1-\gamma) \log p_\text{att} (Y^u|G^u) \bigr\} \label{eq:joint_dec_pft},
\end{align}
where $p_\text{ctc}(Y^u|G^u)$ and $p_\text{att} (Y^u|G^u)$ are obtained with the same encoder output $G^u$.

\subsection{Promoting separation of hidden vectors}
\label{subsec:neg_kl_loss}
A single decoder network is used to output multiple label sequences by independently decoding the multiple hidden vectors generated by the encoder network.
In order for the decoder to generate multiple different label sequences the encoder needs to generate sufficiently differentiated hidden vector sequences for each speaker. 
We propose to encourage this contrast among hidden vectors by introducing in the objective function a new term based on the negative symmetric Kullback-Leibler (KL) divergence.
In the particular case of two-speaker mixtures, we consider the following additional loss function:
\begin{align}
    \mathcal{L}_{\mathrm{KL}} = -\eta \sum_l \bigl\{ &\mathrm{KL}(\bar{G}^1(l) \ || \ \bar{G}^2(l)) \nonumber \\
    &+ \mathrm{KL}(\bar{G}^2(l) \ || \ \bar{G}^1(l)) \bigr\}, \label{eq:kl_loss}
\end{align}
where $\eta$ is a small constant value, and ${\bar{G}^u = ( \mathrm{softmax} (G^u(l)) \ |\  l = 1, \ldots, L ) }$ is obtained from the hidden vector sequence $G^u$ at the output of the recognition encoder ${\mathrm{ Encoder_{Rec}}}$ as in Fig.~\ref{fig:end-to-end-pft} by applying an additional frame-wise softmax operation in order to obtain a quantity amenable to a probability distribution.

\subsection{Split of hidden vector for multiple hypotheses}
\label{sec:nnarch}
Since the network maps acoustic features to label sequences directly, we consider various architectures to perform implicit separation and recognition effectively.
As a baseline system, we use the concatenation of a VGG-motivated CNN network~\cite{simonyan2014very} (referred to as VGG) and a bi-directional long short-term memory (BLSTM) network as the encoder network.
For the splitting point in the hidden vector computation, we consider two architectural variations as follows:
\begin{itemize}[itemindent=0mm,leftmargin=4mm]
\item Split by BLSTM: The hidden vector is split at the level of the BLSTM network.
1) the VGG network generates a single hidden vector $H$;
2) $H$ is fed into $S$ independent BLSTMs whose parameters are not shared with each other;
3) the output of each independent BLSTM $H^u, u\!=\!1,\dots,S,$ is further separately fed into a unique BLSTM, the same for all outputs.
Each step corresponds to Eqs.~(\ref{eq:enc_mix}), (\ref{eq:enc_sd}), and (\ref{eq:enc_rec}).
\item Split by VGG: 
The hidden vector is split at the level of the VGG network. The number of filters at the last convolution layer is multiplied by the number of mixtures $S$ in order to split the output into $S$ hidden vectors (as in Eq.~(\ref{eq:enc_sd})). The layers prior to the last VGG layer correspond to the network in Eq.~(\ref{eq:enc_mix}), while the subsequent BLSTM layers implement the network in (\ref{eq:enc_rec}).
\end{itemize}

\section{Experiments}
\subsection{Experimental setup}
We used English and Japanese speech corpora, WSJ (Wall street journal)~\cite{wsj1, garofalo2007csr} and CSJ (Corpus of spontaneous Japanese)~\cite{maekawa2003corpus}.
To show the effectiveness of the proposed models, we generated mixed speech signals from these corpora to simulate single-channel overlapped multi-speaker recording, and evaluated the recognition performance using the mixed speech data.
For WSJ, we used WSJ1 SI284 for training, Dev93 for development, and Eval92 for evaluation.
For CSJ, we followed the Kaldi recipe~\cite{moriya2015kaldi} and used the full set of academic and simulated presentations for training, and the standard test sets 1, 2, and 3 for evaluation.

We created new corpora by mixing two utterances with different speakers sampled from existing corpora.
The detailed algorithm is presented in Section~\ref{sup:setup} of the supplementary material.
The sampled pairs of two utterances are mixed at various signal-to-noise ratios (SNR) between 0~dB and 5~dB with a random starting point for the overlap.
Duration of original unmixed and generated mixed corpora are summarized in Table~\ref{tab:corpora}.
\begin{table}[t]
\caption{Duration (hours) of unmixed and mixed corpora. The mixed corpora are generated by Algorithm~\ref{algo:data_gen} in Section~\ref{sup:setup} of the supplementary material, using the training, development, and evaluation set respectively.}
\label{tab:corpora}
\vskip -0.1in
\centering
\begin{small}
\begin{sc}
\begin{tabular}{l|ccc}
\hline
 & Train & Dev. & Eval \\
\hline
WSJ (unmixed) & \phantom{0}81.5 & 1.1 & 0.7 \\
WSJ (mixed) & \phantom{0}98.5 & 1.3 & 0.8 \\
\hline
CSJ (unmixed) & 583.8 & 6.6 & 5.2 \\
CSJ (mixed) & 826.9 & 9.1 & 7.5 \\
\hline
\end{tabular}
\end{sc}
\end{small}
\vskip -0.1in
\end{table}

\subsubsection{Network architecture}
As input feature, we used 80-dimensional log Mel filterbank coefficients
with pitch features and their delta and delta delta features ($83\times3 = 249$-dimension)
extracted using Kaldi tools~\cite{Povey_ASRU2011}.
The input feature is normalized to zero mean and unit variance.
As a baseline system, we used a stack of a 6-layer VGG network and a 7-layer BLSTM as the encoder network.
Each BLSTM layer has 320 cells in each direction, and is followed by a linear projection layer with 320 units to combine the forward and backward LSTM outputs.
The decoder network has an 1-layer LSTM with 320 cells.
As described in Section~\ref{sec:nnarch}, we adopted two types of encoder architectures for multi-speaker speech recognition.
The network architectures are summarized in Table~\ref{tab:submodules}.
The split-by-VGG network had speaker differentiating encoders with a convolution layer (and the following maxpooling layer).
The split-by-BLSTM network had speaker differentiating encoders with two BLSTM layers.
The architectures were adjusted to have the same number of layers.
We used characters as output labels.
The number of characters for WSJ was set to 49 including alphabets and special tokens (e.g., characters for space and unknown).
The number of characters for CSJ was set to 3,315 including Japanese Kanji/Hiragana/Katakana characters and special tokens.
\begin{table}[t]
\caption{Network architectures for the encoder network. The number of layers is indicated in parentheses. $\mathrm{Encoder_{Mix}}$, $\mathrm{Encoder_{SD}^u}$, and $\mathrm{Encoder_{Rec}}$ correspond to Eqs.~(\ref{eq:enc_mix}),~(\ref{eq:enc_sd}), and~(\ref{eq:enc_rec}).}
\label{tab:submodules}
\vskip -0.1in
\centering
\begin{small}
\begin{sc}
\begin{tabular}{l|ccc}
\hline
Split by & $\mathrm{Encoder_{Mix}}$ & $\mathrm{Encoder_{SD}^u}$ & $\mathrm{Encoder_{Rec}}$ \\
\hline
no & VGG (6) & --- & BLSTM (7) \\
VGG & VGG (4) & VGG (2) & BLSTM (7) \\
BLSTM & VGG (6) & BLSTM (2) & BLSTM (5) \\
\hline
\end{tabular}
\end{sc}
\end{small}
\vskip -0.2in
\end{table}

\subsubsection{Optimization}
The network was initialized randomly from uniform distribution in the range -0.1 to 0.1.
We used the AdaDelta algorithm~\cite{zeiler2012adadelta} with gradient clipping~\cite{pascanu2013difficulty} for optimization.
We initialized the AdaDelta hyperparameters as $\rho = 0.95$ and $\epsilon = 1^{-8}$.
$\epsilon$ is decayed by half when the loss on the development set degrades.
The networks were implemented with Chainer~\cite{tokui2015chainer} and ChainerMN~\cite{akiba2017chainermn}.
The optimization of the networks was done by synchronous data parallelism with 4 GPUs for WSJ and 8 GPUs for CSJ.

The networks were first trained on single-speaker speech, and then retrained with mixed speech.
When training on unmixed speech, only one side of the network only (with a single speaker differentiating encoder) is optimized to output the label sequence of the single speaker. Note that only character labels are used, and there is no need for clean source reference corresponding to the mixed speech.
When moving to mixed speech, the other speaker-differentiating encoders are initialized using the already trained one by copying the parameters with random perturbation,
$w' = w \times (1 + \mathrm{Uniform}(-0.1, 0.1))$ for each parameter $w$.
The interpolation value $\lambda$ for the multiple objectives in Eqs.~(\ref{eq:single_spk_mtl}) and  (\ref{eq:multi_spk_mtl}) was set to $0.1$ for WSJ and to $0.5$ for CSJ.
Lastly, the model is retrained with the additional negative KL divergence loss in Eq.~(\ref{eq:kl_loss})  with $\eta = 0.1$.

\subsubsection{Decoding}
In the inference stage, we combined a pre-trained RNNLM (recurrent neural network language model) in parallel with the CTC and decoder network. Their label probabilities were linearly combined in the log domain during beam search to find the most likely hypothesis.
For the WSJ task, we used both character and word level RNNLMs~\cite{hori2017multilevel}, where the character model had a 1-layer LSTM with 800 cells and an output layer for 49 characters.
The word model had a 1-layer LSTM with 1000 cells and an output layer for 20,000 words, i.e., the vocabulary size was 20,000.
Both models were trained with the WSJ text corpus.
For the CSJ task, we used a character level RNNLM~\cite{hori2017interspeech}, which had a 1-layer LSTM with 1000 cells and an output layer for 3,315 characters.
The model parameters were trained with the transcript of the training set in CSJ.
We added language model probabilities with an interpolation factor of 0.6 for character-level RNNLM and 1.2 for word-level RNNLM.

The beam width for decoding was set to 20 in all the experiments.
Interpolation $\gamma$ in Eqs. (\ref{eq:joint_dec}) and (\ref{eq:joint_dec_pft}) was set to 0.4 for WSJ and 0.5 for CSJ.

\subsection{Results}
\subsubsection{Evaluation of unmixed speech}
First, we examined the performance of the baseline joint CTC/attention-based encoder-decoder network with the original unmixed speech data.
Table~\ref{table:unmix_wo_pit} shows the character error rates (CERs), %
where the baseline model showed 2.6\% on WSJ and 7.8\% on CSJ.
Since the model was trained and evaluated with unmixed speech data, these CERs are considered lower bounds for the CERs in the succeeding experiments with mixed speech data.
\begin{table}[t]
\caption{Evaluation of unmixed speech without multi-speaker training.}
\label{table:unmix_wo_pit}
\vskip -0.1in
\centering
\begin{small}
\begin{sc}
\begin{tabular}{c|ccc}
\hline
Task & Avg. \\
\hline
WSJ & 2.6 \\
CSJ & 7.8 \\
\hline
\end{tabular}
\end{sc}
\end{small}
\vskip -0.2in
\end{table}

\subsubsection{Evaluation of mixed speech}
Table~\ref{table:mixed_wsj} shows the CERs of the generated mixed speech from the WSJ corpus.
The first column indicates the position of split as mentioned in Section~\ref{sec:nnarch}. The second, third and forth columns indicate CERs of the high energy speaker ({\sc High E. spk.}), the low energy speaker ({\sc Low E. spk.}), and the average ({\sc Avg}.), respectively.
The baseline model has very high CERs because it was trained as a single-speaker speech recognizer without permutation-free training, and it can only output one hypothesis for each mixed speech. In this case, the CERs were calculated by duplicating the generated hypothesis and comparing the duplicated hypotheses with the corresponding references.
The proposed models, i.e., split-by-VGG and split-by-BLSTM networks, obtained significantly lower CERs than the baseline CERs, the split-by-BLSTM model in particular achieving 14.0\% CER. This is an 83.1\% relative reduction from the baseline model.
The CER was further reduced to 13.7\% by retraining the split-by-BLSTM model with the negative KL loss, a 2.1\% relative reduction from the network without retraining.
This result implies that the proposed negative KL loss provides better separation by actively improving the contrast between the hidden vectors of each speaker. 
Examples of recognition results are shown in Section~\ref{sup:recosample} of the supplementary material.
Finally, we profiled the computation time for the permutations based on the decoder network and on CTC.
Permutation based on CTC was 16.3 times faster than that based on the decoder network, in terms of the time required to determine the best match permutation given the encoder network's output in Eq.~(\ref{eq:enc_rec}).

Table~\ref{table:mixed_csj} shows the CERs for the mixed speech from the CSJ corpus.
Similarly to the WSJ experiments, our proposed model significantly reduced the CER from the baseline, where the average CER was 14.9\% and the reduction ratio from the baseline was 83.9\%.

\begin{table}[t]
\caption{CER (\%) of mixed speech for WSJ.}
\label{table:mixed_wsj}
\vskip -0.1in
\centering
\begin{small}
\begin{sc}
\begin{tabular}{l|ccc}
\hline
Split & High E. spk.\!\! & Low E. spk.\!\! & Avg. \\
\hline
no (baseline) & 86.4 & 79.5 & 83.0 \\
VGG & 17.4 & 15.6 & 16.5 \\
BLSTM & 14.6 & {\bf 13.3} & 14.0 \\
+ KL loss & {\bf 14.0} & {\bf 13.3} & {\bf 13.7} \\
\hline
\end{tabular}
\end{sc}
\end{small}
\vskip -0.1in
\end{table}

\begin{table}[t]
\caption{CER (\%) of mixed speech for CSJ.}
\label{table:mixed_csj}
\vskip -0.1in
\centering
\begin{small}
\begin{sc}
\begin{tabular}{l|ccc}
\hline
Split & High E. spk.\!\! & Low E. spk.\!\! & Avg. \\
\hline
no (baseline) & 93.3 & 92.1 & 92.7 \\
BLSTM & {\bf 11.0} & {\bf 18.8} & {\bf 14.9} \\
\hline
\end{tabular}
\end{sc}
\end{small}
\vskip -0.2in
\end{table}

\begin{figure*}[t]
\centering
\includegraphics[width=11.5cm]{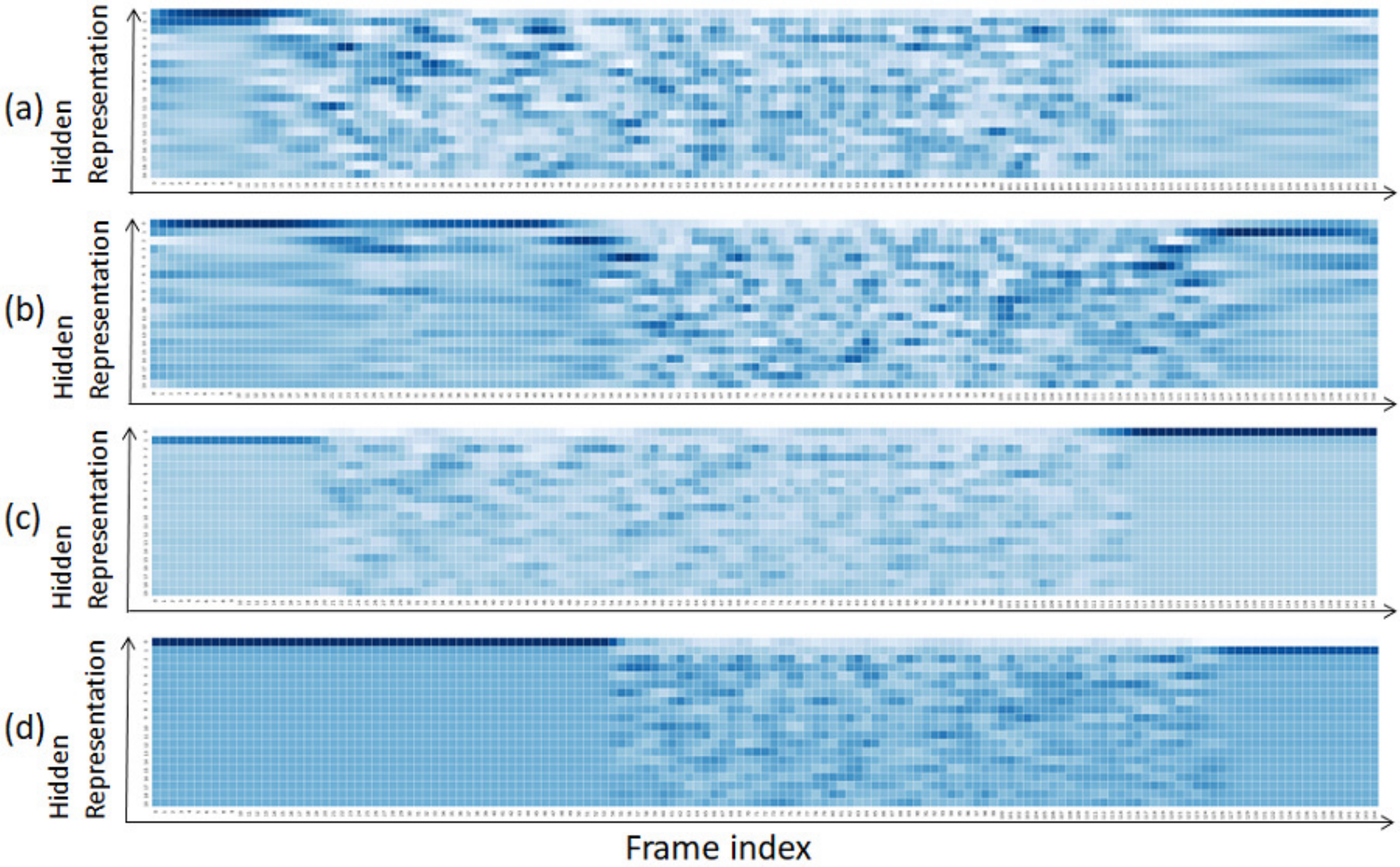}
\vskip -0.1in
\caption{Visualization of the two hidden vector sequences at the output of the split-by-BLSTM encoder on a two-speaker mixture.
(a,b): Generated by the model without the negative KL loss.
(c,d): Generated by the model with the negative KL loss.}
\label{fig:heatmap}
\vskip -0.1in
\end{figure*}
\subsubsection{Visualization of hidden vectors}
We show a visualization of the encoder networks outputs in Fig.~\ref{fig:heatmap} to illustrate the effect of the negative KL loss function.
Principal component analysis (PCA) was applied to the hidden vectors on the vertical axis.
Figures~\ref{fig:heatmap}(a) and~\ref{fig:heatmap}(b) show the hidden vectors generated by the split-by-BLSTM model without the negative KL divergence loss for an example mixture of two speakers.
We can observe different activation patterns showing that the hidden vectors were successfully separated to the individual utterances in the mixed speech, although some activity from one speaker can be seen as leaking into the other.
Figures~\ref{fig:heatmap}(c) and~\ref{fig:heatmap}(d) show the hidden vectors generated after retraining with the negative KL divergence loss.
We can more clearly observe the different patterns and boundaries of activation and deactivation of hidden vectors.
The negative KL loss appears to regularize the separation process, and even seems to help in finding the end-points of the speech.

\subsubsection{Comparison with earlier work}
We first compared the recognition performance with a hybrid (non end-to-end) system including DPCL-based speech separation and a Kaldi-based ASR system. It was evaluated under the same evaluation data and metric as in~\cite{Isik2016} based on the WSJ corpus. However, there are differences in the size of training data and the options in decoding step. Therefore, it is not a fully matched condition. Results are shown in Table~\ref{table:dpcl}. The word error rate (WER) reported in~\cite{Isik2016} is 30.8\%, which was obtained with jointly trained DPCL and second-stage speech enhancement networks. The proposed end-to-end ASR gives an 8.4\% relative reduction in WER even though our model does not require any explicit frame-level labels such as phonetic alignment, or clean signal reference, and does not use a phonetic lexicon for training. Although this is an unfair comparison, our purely end-to-end system outperformed a hybrid system for multi-speaker speech recognition.

Next, we compared our method with an end-to-end explicit separation and recognition network~\cite{Settle2018ICASSP04}. We retrained our model previously trained on our WSJ-based corpus using the training data generated by Settle et al.~\shortcite{Settle2018ICASSP04}, because the direct optimization from scratch on their data caused poor recognition performance due to data size. 
Other experimental conditions are shared with the earlier work. Interestingly, our method showed comparable performance to the end-to-end explicit separation and recognition network, without having to pre-train using clean signal training references.  
It remains to be seen if this parity of performance holds in other tasks and conditions.  

\begin{table}[t]
\caption{Comparison with conventional approaches}
\label{table:dpcl}
\vskip -0.1in
\centering
\begin{small}
\begin{sc}
\begin{tabular}{lc}
\hline
Method & WER (\%) \\
\hline
DPCL + ASR~\cite{Isik2016} & 30.8 \\
{\bf Proposed end-to-end ASR} & {\bf 28.2} \\
\hline
Method & CER (\%) \\
\hline
\shortstack{End-to-end DPCL + ASR (char LM) \\ \cite{Settle2018ICASSP04}} & {\bf 13.2} \\
{\bf Proposed end-to-end ASR (char LM)} & 14.0 \\
\hline
\end{tabular}
\end{sc}
\end{small}
\vskip -0.2in
\end{table}

\section{Related work}
Several previous works have considered an explicit two-step procedure~\cite{john2016deep, Isik2016, dong2017permutation, chen2017deep, chen2018progressive}.
In contrast with our work which uses a single objective function for ASR,
they introduced an objective function to guide the separation of mixed speech.

Qian et al.~\shortcite{Qian2017single} trained a multi-speaker speech recognizer using permutation-free training without explicit objective function for separation.
In contrast with our work which uses an end-to-end architecture, 
their objective function relies on a senone posterior probability obtained by aligning unmixed speech and text using a model trained as a recognizer for single-speaker speech.
Compared with~\cite{Qian2017single}, our method directly maps a speech mixture to multiple character sequences 
and eliminates the need for the corresponding isolated speech sources for training.

\section{Conclusions}
In this paper, we proposed an end-to-end multi-speaker speech recognizer based on permutation-free training and a
new objective function promoting the separation of hidden vectors in order to generate multiple hypotheses.
In an encoder-decoder network framework, teacher forcing at the decoder network under multiple references increases computational cost if implemented naively.
We avoided this problem by employing a joint CTC/attention-based encoder-decoder network.

Experimental results showed that the model is able to directly convert an input speech mixture %
into multiple label sequences
under the end-to-end framework without the need for any explicit intermediate representation including phonetic alignment information or pairwise unmixed speech.
We also compared our model with a method based on explicit separation using deep clustering, and showed comparable result.
Future work includes data collection and evaluation in a real world scenario since the data used in our experiments are simulated mixed speech, which is already extremely challenging but still leaves some acoustic aspects, such as Lombard effects and real room impulse responses, that need to be alleviated for further performance improvement.
In addition, further study is required in terms of increasing the number of speakers that can be simultaneously recognized, and further comparison with the separation-based approach.

\bibliography{acl2018}

\begin{thebibliography}{26}
\expandafter\ifx\csname natexlab\endcsname\relax\def\natexlab#1{#1}\fi

\bibitem[{Akiba et~al.(2017)Akiba, Fukuda, and Suzuki}]{akiba2017chainermn}
Takuya Akiba, Keisuke Fukuda, and Shuji Suzuki. 2017.
\newblock {ChainerMN: Scalable Distributed Deep Learning Framework}.
\newblock In \emph{Proceedings of Workshop on ML Systems in The Thirty-first
  Annual Conference on Neural Information Processing Systems (NIPS)}.

\bibitem[{Bahdanau et~al.(2016)Bahdanau, Chorowski, Serdyuk, Brakel, and
  Bengio}]{bahdanau2016end}
Dzmitry Bahdanau, Jan Chorowski, Dmitriy Serdyuk, Philemon Brakel, and Yoshua
  Bengio. 2016.
\newblock End-to-end attention-based large vocabulary speech recognition.
\newblock In \emph{IEEE International Conference on Acoustics, Speech and
  Signal Processing (ICASSP)}, pages 4945--4949.

\bibitem[{Chen et~al.(2018)Chen, Droppo, Li, and Xiong}]{chen2018progressive}
Zhehuai Chen, Jasha Droppo, Jinyu Li, and Wayne Xiong. 2018.
\newblock Progressive joint modeling in unsupervised single-channel overlapped
  speech recognition.
\newblock \emph{IEEE/ACM Transactions on Audio, Speech, and Language
  Processing}, 26(1):184--196.

\bibitem[{Chen et~al.(2017)Chen, Luo, and Mesgarani}]{chen2017deep}
Zhuo Chen, Yi~Luo, and Nima Mesgarani. 2017.
\newblock Deep attractor network for single-microphone speaker separation.
\newblock In \emph{IEEE International Conference on Acoustics, Speech and
  Signal Processing (ICASSP)}, pages 246--250.

\bibitem[{Chorowski et~al.(2015)Chorowski, Bahdanau, Serdyuk, Cho, and
  Bengio}]{chorowski2015attention}
Jan~K Chorowski, Dzmitry Bahdanau, Dmitriy Serdyuk, Kyunghyun Cho, and Yoshua
  Bengio. 2015.
\newblock Attention-based models for speech recognition.
\newblock In \emph{Advances in Neural Information Processing Systems (NIPS)},
  pages 577--585.

\bibitem[{Consortium(1994)}]{wsj1}
Linguistic~Data Consortium. 1994.
\newblock {CSR}-{II} (wsj1) complete.
\newblock \emph{Linguistic Data Consortium, Philadelphia}, LDC94S13A.

\bibitem[{Cooke et~al.(2009)Cooke, Hershey, and Rennie}]{Cooke2009monaural}
Martin Cooke, John~R Hershey, and Steven~J Rennie. 2009.
\newblock Monaural speech separation and recognition challenge.
\newblock \emph{Computer Speech and Language}, 24(1):1--15.

\bibitem[{Garofalo et~al.(2007)Garofalo, Graff, Paul, and
  Pallett}]{garofalo2007csr}
John Garofalo, David Graff, Doug Paul, and David Pallett. 2007.
\newblock {CSR}-{I} (wsj0) complete.
\newblock \emph{Linguistic Data Consortium, Philadelphia}, LDC93S6A.

\bibitem[{Graves et~al.(2006)Graves, Fern{\'a}ndez, Gomez, and
  Schmidhuber}]{graves2006connectionist}
Alex Graves, Santiago Fern{\'a}ndez, Faustino Gomez, and J{\"u}rgen
  Schmidhuber. 2006.
\newblock Connectionist temporal classification: labelling unsegmented sequence
  data with recurrent neural networks.
\newblock In \emph{International Conference on Machine learning (ICML)}, pages
  369--376.

\bibitem[{Hershey et~al.(2016)Hershey, Chen, Le~Roux, and
  Watanabe}]{john2016deep}
John~R Hershey, Zhuo Chen, Jonathan Le~Roux, and Shinji Watanabe. 2016.
\newblock Deep clustering: Discriminative embeddings for segmentation and
  separation.
\newblock In \emph{IEEE International Conference on Acoustics, Speech and
  Signal Processing (ICASSP)}, pages 31--35.

\bibitem[{Hori et~al.(2017{\natexlab{a}})Hori, Watanabe, and
  Hershey}]{hori-ACL-2017}
Takaaki Hori, Shinji Watanabe, and John~R Hershey. 2017{\natexlab{a}}.
\newblock Joint {CTC}/attention decoding for end-to-end speech recognition.
\newblock In \emph{Proceedings of the 55th Annual Meeting of the Association
  for Computational Linguistics (ACL): Human Language Technologies: long
  papers}.

\bibitem[{Hori et~al.(2017{\natexlab{b}})Hori, Watanabe, and
  Hershey}]{hori2017multilevel}
Takaaki Hori, Shinji Watanabe, and John~R Hershey. 2017{\natexlab{b}}.
\newblock Multi-level language modeling and decoding for open vocabulary
  end-to-end speech recognition.
\newblock In \emph{IEEE Workshop on Automatic Speech Recognition and
  Understanding (ASRU)}.

\bibitem[{Hori et~al.(2017{\natexlab{c}})Hori, Watanabe, Zhang, and
  William}]{hori2017interspeech}
Takaaki Hori, Shinji Watanabe, Yu~Zhang, and Chan William. 2017{\natexlab{c}}.
\newblock Advances in joint {CTC}-{A}ttention based end-to-end speech
  recognition with a deep {CNN} encoder and {RNN}-{LM}.
\newblock In \emph{Interspeech}, pages 949--953.

\bibitem[{Isik et~al.(2016)Isik, {Le Roux}, Chen, Watanabe, and
  Hershey}]{Isik2016}
Yusuf Isik, Jonathan {Le Roux}, Zhuo Chen, Shinji Watanabe, and John~R.
  Hershey. 2016.
\newblock Single-channel multi-speaker separation using deep clustering.
\newblock In \emph{Proc. Interspeech}, pages 545--549.

\bibitem[{Kim et~al.(2017)Kim, Hori, and Watanabe}]{kim2016joint_icassp2017}
Suyoun Kim, Takaaki Hori, and Shinji Watanabe. 2017.
\newblock Joint {CTC}-attention based end-to-end speech recognition using
  multi-task learning.
\newblock In \emph{IEEE International Conference on Acoustics, Speech and
  Signal Processing (ICASSP)}, pages 4835--4839.

\bibitem[{Kolb{\ae}k et~al.(2017)Kolb{\ae}k, Yu, Tan, and
  Jensen}]{kolbaek2017multitalker}
Morten Kolb{\ae}k, Dong Yu, Zheng-Hua Tan, and Jesper Jensen. 2017.
\newblock Multitalker speech separation with utterance-level permutation
  invariant training of deep recurrent neural networks.
\newblock \emph{IEEE/ACM Transactions on Audio, Speech, and Language
  Processing}, 25(10):1901--1913.

\bibitem[{Maekawa(2003)}]{maekawa2003corpus}
Kikuo Maekawa. 2003.
\newblock Corpus of {Spontaneous} {Japanese}: Its design and evaluation.
\newblock In \emph{ISCA \& IEEE Workshop on Spontaneous Speech Processing and
  Recognition}.

\bibitem[{Moriya et~al.(2015)Moriya, Shinozaki, and Watanabe}]{moriya2015kaldi}
Takafumi Moriya, Takahiro Shinozaki, and Shinji Watanabe. 2015.
\newblock Kaldi recipe for {Japanese} spontaneous speech recognition and its
  evaluation.
\newblock In \emph{Autumn Meeting of ASJ}, 3-Q-7.

\bibitem[{Pascanu et~al.(2013)Pascanu, Mikolov, and
  Bengio}]{pascanu2013difficulty}
Razvan Pascanu, Tomas Mikolov, and Yoshua Bengio. 2013.
\newblock On the difficulty of training recurrent neural networks.
\newblock \emph{International Conference on Machine Learning (ICML)}, pages
  1310--1318.

\bibitem[{Povey et~al.(2011)Povey, Ghoshal, Boulianne, Burget, Glembek, Goel,
  Hannemann, Motlicek, Qian, Schwarz, Silovsky, Stemmer, and
  Vesely}]{Povey_ASRU2011}
Daniel Povey, Arnab Ghoshal, Gilles Boulianne, Lukas Burget, Ondrej Glembek,
  Nagendra Goel, Mirko Hannemann, Petr Motlicek, Yanmin Qian, Petr Schwarz, Jan
  Silovsky, Georg Stemmer, and Karel Vesely. 2011.
\newblock The kaldi speech recognition toolkit.
\newblock In \emph{IEEE Workshop on Automatic Speech Recognition and
  Understanding (ASRU)}.

\bibitem[{Qian et~al.(2017)Qian, Chang, and Yu}]{Qian2017single}
Yanmin Qian, Xuankai Chang, and Dong Yu. 2017.
\newblock Single-channel multi-talker speech recognition with permutation
  invariant training.
\newblock \emph{arXiv preprint arXiv:1707.06527}.

\bibitem[{Settle et~al.(2018)Settle, {Le Roux}, Hori, Watanabe, and
  Hershey}]{Settle2018ICASSP04}
Shane Settle, Jonathan {Le Roux}, Takaaki Hori, Shinji Watanabe, and John~R.
  Hershey. 2018.
\newblock End-to-end multi-speaker speech recognition.
\newblock In \emph{IEEE International Conference on Acoustics, Speech and
  Signal Processing (ICASSP)}, pages 4819--4823.

\bibitem[{Simonyan and Zisserman(2014)}]{simonyan2014very}
Karen Simonyan and Andrew Zisserman. 2014.
\newblock Very deep convolutional networks for large-scale image recognition.
\newblock \emph{arXiv preprint arXiv:1409.1556}.

\bibitem[{Tokui et~al.(2015)Tokui, Oono, Hido, and Clayton}]{tokui2015chainer}
Seiya Tokui, Kenta Oono, Shohei Hido, and Justin Clayton. 2015.
\newblock Chainer: a next-generation open source framework for deep learning.
\newblock In \emph{Proceedings of Workshop on Machine Learning Systems
  (LearningSys) in NIPS}.

\bibitem[{Yu et~al.(2017)Yu, Kolbæk, Tan, and Jensen}]{dong2017permutation}
Dong Yu, Morten Kolbæk, Zheng-Hua Tan, and Jesper Jensen. 2017.
\newblock Permutation invariant training of deep models for speaker-independent
  multi-talker speech separation.
\newblock In \emph{IEEE International Conference on Acoustics, Speech and
  Signal Processing (ICASSP)}, pages 241--245.

\bibitem[{Zeiler(2012)}]{zeiler2012adadelta}
Matthew~D Zeiler. 2012.
\newblock {ADADELTA}: an adaptive learning rate method.
\newblock \emph{arXiv preprint arXiv:1212.5701}.

\end{thebibliography}
\bibliographystyle{acl_natbib}

\appendix

\clearpage
\section{Architecture of the encoder-decoder network}
\label{sup:encdec}
In this section, we describe the details of the baseline encoder-decoder network which is further extended for permutation-free training.
The encoder network consists of a VGG network and bi-directional long short-term memory (BLSTM) layers.
The VGG network has the following 6-layer CNN architecture at the bottom of the encoder network:
\begin{description}
    \setlength{\itemsep}{-3pt}
    \item {\normalsize Convolution (\# in = 3, \# out = 64, filter = 3$\times$3)}
    \item {\normalsize Convolution (\# in = 64, \# out = 64, filter = 3$\times$3)}
    \item {\normalsize MaxPooling (patch = 2$\times$2, stride = 2$\times$2)}
    \item {\normalsize Convolution (\# in = 64, \# out = 128, filter = 3$\times$3)}
    \item {\normalsize Convolution (\# in=128, \# out=128, filter=3$\times$3)}
    \item {\normalsize MaxPooling (patch = 2$\times$2, stride = 2$\times$2)}
\end{description}
The first 3 channels are static, delta, and delta delta features.
Multiple BLSTM layers with projection layer $\mathrm{Lin}(\cdot)$ are stacked after the VGG network.
We defined one BLSTM layer as the concatenation of a forward LSTM $\overrightarrow{\mathrm{LSTM}(\cdot)}$ and a backward LSTM $\overleftarrow{\mathrm{LSTM}(\cdot)}$:
\begin{align}
    \overrightarrow{H} &= \overrightarrow{\mathrm{LSTM}}(\cdot) \\
    \overleftarrow{H} &= \overleftarrow{\mathrm{LSTM}}(\cdot) \\
    H &= [\mathrm{Lin}(\overrightarrow{H}); \mathrm{Lin}(\overleftarrow{H})],
\end{align}
When the VGG network and the multiple BLSTM layers are represented as $\mathrm{VGG}(\cdot)$ and $\mathrm{BLSTM}(\cdot)$, 
the encoder network in Eq.~(\ref{eq:enc}) maps the input feature vector $O$ to internal representation $H$ as follows:
\begin{align}
    H = \mathrm{Encoder}(O) = \mathrm{BLSTM}(\mathrm{VGG}(O))
\end{align}

The decoder network sequentially generates the $n$-th label $y_n$ by taking the context vector $c_n$ and the label history $y_{1:n-1}$:
\begin{align}
    y_{n} & \sim \mathrm{Decoder}(c_{n}, y_{n-1}).  \label{supeq:dec}
\end{align}
The context vector is calculated in an location based attention mechanism~\cite{chorowski2015attention}
which weights and sums the $C$-dimensional sequence of representation $H = ( h_l \in \mathbb{R}^C | l = 1, \dots, L)$ with attention weight $a_{n,l}$:
\begin{align}
    c_{n} &= \mathrm{Attention}(a_{n-1}, e_{n}, H), \label{supeq:att} \\
    &\triangleq \sum_{l=1}^{L} a_{n,l} h_{l}.
\end{align}
The location based attention mechanism defines the weights $a_{n,l}$ as follows:
\begin{align}
    a_{n,l} &= \frac{\exp(\alpha k_{n,l})}{\sum_{l=1}^{L} \exp(\alpha k_{n,l})}, \\
    k_{n,l} &= w^{\mathrm{T}} \mathrm{tanh}(
    V^{\text{E}} e_{n-1} + 
    V^{\text{H}} h_{l} + 
    V^{\text{F}} f_{n,l} + 
    b), \label{supeq:att_mid} \\
    f_{n} &= F * a_{n-1}, \label{supeq:att_begin}
\end{align}
where $w, V^E, V^H, V^F, b, F$ are tunable parameters, $\alpha$ is a constant value called inverse temperature,
and $*$ is the convolution operation.
We used $10$ convolution filters of width $200$, and set $\alpha$ to 2.
The introduction of $f_n$ makes the attention mechanism take into account the previous alignment information.
The hidden state $e$ is updated recursively by an updating LSTM function:
\begin{align}
    e_{n} &= \mathrm{Update}(e_{n-1}, c_{n-1}, y_{n-1}), \\
    &\triangleq \mathrm{LSTM}( \nonumber \\ 
    &\mathrm{Lin}(e_{n-1}) + \mathrm{Lin}(c_{n-1}) + \mathrm{Emb}(y_{n-1})),
\end{align}
where $\mathrm{Emb}(\cdot)$ is an embedding function.

\begin{algorithm}[tb]
    \caption{Generation of multi speaker speech dataset}
    \label{algo:data_gen}
    \begin{algorithmic}
    \STATE{$n_{\text{reuse}} \Leftarrow$ maximum number of times same utterance can be used.}
    \STATE{$U \Leftarrow$ utterance set of the corpora.}
    \STATE{$C_k \Leftarrow$ $n_{\text{reuse}}$ for each utterance $U_k \in U$}
    \FOR{$U_{k} \in U$}
        \STATE{$P(U_{k})$ = $C_{k}$ / $\sum_l C_{l}$}
    \ENDFOR
    \FOR{$U_{i}$ in $U$}
        \STATE{Sample utterance $U_{j}$ from $P(U)$ while ensuring speakers of $U_{i}$ and $U_{j}$ are different.}
        \STATE{Mix utterances $U_i$ and $U_j$}
        \IF{$C_{j} > 0$}
            \STATE{$C_{j}$ = $C_{j} - 1$}
            \FOR{$U_{k} \Leftarrow U$}
                \STATE{$P(U_{k})$ = $C_{k}$ / $\sum_l C_{l}$}
            \ENDFOR
        \ENDIF
    \ENDFOR
    \end{algorithmic}
\end{algorithm}

\begin{table*}[tb]
    \centering 
    \caption{Examples of recognition results. Errors are emphasized as capital letter. ``$\_$'' is a space character, and a special token ``*'' is inserted to pad deletion errors.}
    \label{tb:ex_trans}
    \begin{tabular}{p{15cm}}\hline
    {\bf \footnotesize{ (1) Model w/ permutation-free training (CER of HYP1: 12.8\%, HYP2: 0.9\%)}} \\
    {\footnotesize {\bf HYP1:} t h e \_ s h u t t l e \_ * * * I S \_ I N \_ t h e \_ f i r s t \_ t H E \_ l i f E \_ o * f \_ s i n c e \_ t h e \_ n i n e t e e n \_ e i g h t y \_ s i x \_ c h a l l e n g e r \_ e x p l o s i o n} \\
    {\footnotesize {\bf REF1}: t h e \_ s h u t t l e \_ W O U L D \_ B E \_ t h e \_ f i r s t \_ t * O \_ l i f T \_ o F f \_ s i n c e \_ t h e \_ n i n e t e e n \_ e i g h t y \_ s i x \_ c h a l l e n g e r \_ e x p l o s i o n} \\
    {\footnotesize {\bf HYP2:} t h e \_ e x p a n d e d \_ r e c a l l \_ w a s \_ d i s c l o s e d \_ a t \_ a \_ m e e t i n g \_ w i t h \_ n . \_ r . \_ c . \_ o f f i c i a l s \_ a t \_ a n \_ a g e n c y \_ o f f i c e \_ o u t s i d e \_ c h i c a g o} \\
    {\footnotesize {\bf REF2}: t h e \_ e x p a n d e d \_ r e c a l l \_ w a s \_ d i s c l o s e d \_ a t \_ a \_ m e e t i n g \_ w i t h \_ n . \_ r . \_ c . \_ o f f i c i a l s \_ a t \_ a n \_ a g e n c y \_ o f f i c e \_ o u t s i d e \_ c h i c a g o} \\
    \hline
    {\bf \footnotesize{ (2) Model w/ permutation-free training (CER of HYP1: 91.7\%, HYP2: 38.9\%)}} \\
    {\footnotesize {\bf HYP1:} I T \_ W A S \_ L a s t \_ r * A I S e * D \_ * I N \_ J U N E \_ N I N E t E e N \_ e * I G h T Y \_ f I V e \_ T O \_ * \underline{T H I R T Y}} \\
    {\footnotesize {\bf REF1:} * * * * * * * * a s t * r O N O M e R S \_ S A Y \_ T H A T \_ * * * * t H e * \_ e A R T h ' S \_ f A T e \_ I S \_ S E A L E D} \\    
    {\footnotesize {\bf HYP2:} * * * * a N D \_ * * s t * r O N G e R S \_ S A Y \_ T H A T \_ * * * * t H e * \_ e * A R t H \_ f A T e \_ I S \_ t o \_ f o r t y \_ f i v e \_ d o l l a r s \_ f r o m \_ \underline{t h i r t y} \_ f i v e \_ d o l l a r s} \\
    {\footnotesize {\bf REF2:} I T \_ W a * S \_ L A s t \_ r A I S e * D \_ * I N \_ J U N E \_ N I N E t E e N \_ e I G H t Y \_ f I V e * * * \_ t o \_ f o r t y \_ f i v e \_ d o l l a r s \_ f r o m \_ \underline{t h i r t y} \_ f i v e \_ d o l l a r s} \\
    \hline
    \end{tabular}
\end{table*}

\section{Generation of mixed speech}
\label{sup:setup}
Each utterance of the corpus is mixed with a randomly selected utterance with the probability, $P(U_{k})$,
that moderates over-selection of specific utterances.
$P(U_k)$ is calculated in the first for-loop as a uniform probability.
All utterances are used as one side of the mixture, and another side is sampled from the distribution $P(U_k)$ in the second for-loop.
The selected pairs of utterances are mixed at various signal-to-noise ratios (SNR) between 0~dB and 5~dB.
We randomized the starting point of the overlap by padding the shorter utterance with silence whose duration is sampled from the uniform distribution within the length difference between the two utterances.
Therefore, the duration of the mixed utterance is equal to that of the longer utterance among the unmixed speech.
After the generation of the mixed speech, the count of selected utterances $C_j$ is decremented to prevent of over-selection.
All counts $C$ are set to $n_{\mathrm{reuse}}$, and we used $n_{\mathrm{reuse}} = 3$.

\section{Examples of recognition results and error analysis}
\label{sup:recosample}
Table~\ref{tb:ex_trans} shows examples of recognition result.
The first example (1) is one which accounts for a large portion of the evaluation set.
The SNR of the HYP1 is -1.55 db and that of HYP2 is 1.55 dB.
The network generates multiple hypotheses with a few substitution and deletion errors, but without any overlapped and swapped words.
The second example (2) is one which leads to performance reduction.
We can see that the network makes errors when there is a large difference in length between the two sequences.
The word ``thirty'' of HYP2 is injected in HYP1, and there are deletion errors in HYP2.
We added a negative KL divergence loss to ease such kind of errors.
However, there is further room to reduce error by making unshared modules more cooperative.

\end{document}